# Investigation of Nuclear Phase transition with using Landau Theory

H. Fathi, H. Sabri[*]

Department of Nuclear Physics, University of Tabriz, Tabriz 51664, Iran.

[*] Corresponding Author E-mail: h-sabri@tabrizu.ac.ir

Abstract

In this paper, Landau theory for phase transitions is shown to be a useful approach also for quantal system such as atomic nucleus. A detailed analysis of critical exponents of ground state quantum phase transition between $U(5)$ and $O(6)$ limits of interacting boson model is presented.

1. Introduction

Studying the behavior of nuclear matter under extreme conditions of temperature and density, including possible phase transitions, is one of the most interesting subjects in recent years. Drastic changes in properties of physical systems are called phase transitions which these properties have been characterized by order parameters [1-5]. Phase transitions occur as some of parameters, i.e. control parameters, which have constrained system, are varied. Temperature-governed phase transitions in which the control parameter is temperature, $T$, have been known for many years. Landau theory of phase transitions was formulated in the late 1930's as an attempt to develop a general method of analysis for various types of phase transitions in condensed matter physics and especially in crystals [1-2]. It relies on two basic conditions, namely on (a) the assumption that the free energy is an analytic function of a order parameter, and on (b) the fact that the expression for the free energy must obey the symmetries of the system. Condition (a) is further strengthened by expressing the free energy as a Taylor series in the order parameter.

For fluid systems, as we become close to the critical point, some of the quantities of system are related to the temperature as $f(T) \approx (T - T_c)^\beta$ for some exponents of $\beta$. The similar behaviors may be seen not as a function of temperature, but as a function of some other quantities of system, e.g. $f(x) \approx (x)^\beta$. These exponents, $\beta$, are called critical exponents and naturally defined as $\lim_{x \to 0} \frac{\ln f(x)}{\ln(\pm x)}$ [4,5]. Some basic critical exponents in thermodynamics have been employed to describe the evolution of considered systems near the critical points.

2. Quantum Phase Transition (OPT) in Interacting Boson Model (IBM)

In nuclear physics, quantum phase transitions, sometimes called zero temperature or ground-state phase transitions, can be studied most easily with using algebraic techniques that associate with a specific mathematical symmetry with different nuclear shapes. Interacting Boson Model (IBM), as the most popular algebraic model in description of nuclear structures, was proposed in 1975 by Iachello and Arima to describe collective excitations of atomic nuclei [3]. In this model, nucleons in an even-even isotope are divided into an inert core and an even number of valence particles. These particles are then considered as coupled into two kinds of bosons that may carry either a total angular momentum 0 or 2, and are respectively called the *s*- and *d*-bosons. The bilinear operator that may be formed with *s*- and *d*-boson creation and annihilation operators close into the *U(6)* algebra, whose three possible subgroup chain match with the *U(5)*, *SU(3)* and *SO(6)* solution of the Bohr Hamiltonian, i.e. respectively with spherical, axially deformed and γ-unstable shapes. It is of great interest to be able to describe the evolution of considered systems near the critical points. Let's consider a general form of IBM Hamiltonian as [3]

$$H = \varepsilon_d \hat{n}_d + k_0 \hat{Q}^\chi \cdot \hat{Q}^\chi + k_1 \hat{P}^\dagger \cdot \hat{P} + k_2 \hat{L} \cdot \hat{L} + k_3 \hat{T}_3 \cdot \hat{T}_3 + k_4 \hat{T}_4 \cdot \hat{T}_4 \quad , \tag{1}$$

where $\hat{n}_d (= d^\dagger \cdot \tilde{d})$ is the *d* boson number operator, $\hat{Q}^\chi$, i.e. $(s^\dagger \times \tilde{d} + d^\dagger \times \tilde{s})^{(2)}$ explores the quadrupole interaction. Also, other terms of Hamiltonian are

$$\hat{P}^\dagger = \frac{1}{2}(d^\dagger \cdot d^\dagger - s^\dagger \cdot s^\dagger) \quad , \qquad \hat{L} = \sqrt{10}(d^\dagger \times \tilde{d})^{(1)} \tag{2}$$

$$\hat{T}_3 = (d^\dagger \times \tilde{d})^{(3)} \quad , \qquad \hat{T}_4 = (d^\dagger \times \tilde{d})^{(4)}$$

This general Hamiltonian can describe three dynamical symmetry limits with different values of constants, i.e. $\varepsilon_0$, $\chi$ and $k_i'$ s. We must consider a transitional Hamiltonian to describe critical exponents at the critical point of phase transition. To this aim, we propose the following schematic Hamiltonians for $U(5) - O(6)$ transition [6]

$$H_{trans} = x\hat{n}_d + \frac{1-x}{N-1} \hat{P}^\dagger \cdot \hat{P} \quad , \tag{3}$$

Where we have introduced $\varepsilon_d = x$ and $k_1 = (1-x)/(N-1)$. The $U(5)$ limit of IBM is recovered via $x = 1$ and $x = 0$ reproduces $O(6)$ limit. It means, one can describe a continuous, e.g. second-order, shape phase transition by changing $x$ between these two limits. On the other hand, classical limit of transitional Hamiltonian, Eq.(3), is obtained by considering its expectation value in the coherent state [7-8]

$$|N, \alpha_m\rangle = (s^\dagger + \sum_m \alpha_m d_m^\dagger)^N |0\rangle \quad , \tag{4}$$

Where $|0\rangle$ is the boson vacuum state, $s^\dagger$ and $d^\dagger$ are the creation operators of $s$ and $d$ bosons, respectively and $\alpha_m$ can be related to deformation collective parameters, $\alpha_0 = \beta\cos\gamma$, $\alpha_{\pm 1} = 0$ and $\alpha_{\pm 2} = \beta\sin\gamma/\sqrt{2}$. The energy surface which follows from expectation value of transitional Hamiltonian in the coherent state, Eq.(4), is given by

$$E(N, \beta) = N[x\frac{\beta^2}{1+\beta^2} + \frac{1-x}{4}(\frac{1-\beta^2}{1+\beta^2})^2] \quad , \tag{5}$$

Critical point of considered transitional region have obtained via $d^2E(N,\beta)/d\beta^2\big|_{\beta=0} = 0$ [9-12] condition which gives $x_c = 0.5$ in this case. We show the dependence of energy surface on the order parameter, $\beta$, above and below of the critical point of phase transition, $x_c$, in Figure1. In phase transition from $U(5)$, i.e. spherical limit, to $O(6)$, namely, $\gamma$-unstable limit, one sees that, the evolution of energy surface goes from a pure $\beta^2$ to a combination of $\beta^2$ and $\beta^4$ that has a deformed minimum. At the critical point of this transition, energy surface is a pure $\beta^4$. These results interpret that, $d^2E(N,\beta)/d\beta^2\big|_{\beta=0}=0$ condition corresponds approximately to a ''very flat energy surface'', similar to what have happened for the *E(5)* critical point [13], i.e. critical point of $U(5) \leftrightarrow O(6)$ transitional region.

The typical behavior of the order parameter, $\beta$, at a phase transition is shown in Figure2. Here $\beta$ is small and close to $x_c$ and we assume that, energy surfaces can be expanded around $\beta = 0$,

$$E(N,\beta) = [N(\frac{1-x}{4}) + N(2x-1)\beta^2 + \frac{N}{12}(-31x+19)\beta^4 + ....] \quad , \qquad (6)$$

Or can be rewritten in the form

$$E(N,A,B,C;\beta) = A + B\beta^2 + C\beta^4 \quad , \qquad (7)$$

The behavior of $E(N,A,B,C;\beta)$, near the critical point is determined by the signs of the coefficients *B*, *C*. The coefficients *B*, *C* which are functions of $x$, are written as functions of the dimensionless quantity, $\hat{x} = (x - x_{critical})/x_{critical}$,

$$A = N(\frac{1-\hat{x}}{8}) \quad , \qquad B = N\hat{x} \quad , \qquad C = \frac{N}{24}(7 - 31\hat{x}) \qquad (8)$$

where $x_{critical} = 0.5$. Stable systems have $C > 0$ on both sides of $x_{critical} = 0.5$; therefore *C* is represented only as $\frac{7}{24}N$.

The condition for stability is that, $E(N,A,B,C;\beta)$ must be a minimum as a function of $\beta$. From Eq. (7), this condition may be expressed as

$$2B\beta + 4C\beta^3 = 0 \qquad , \qquad (9)$$

where terms above $\beta^4$ are presumed negligible near $x_{critical} = 0.5$ [10]. For $x_{critical} < 0.5$, only real root is $\beta = 0$; on the other hand, for $x_{critical} > 0.5$, the root $\beta = 0$ correspond to a local maximum, and therefore not to equilibrium. The other two roots are found to be $\beta^2 = -\frac{12}{7}\hat{x}$. Consequently, our analysis predicts, the equilibrium order parameter near the critical point should depend on the $\hat{x}$ as

$$\beta = \text{Constant} \times (-\hat{x})^{\frac{1}{2}} \qquad , \qquad (10)$$

which means, critical exponent for order parameter is $1/2$. The behavior of $\beta_0(x)$ is depicted in Figure 3 which is in perfect agreement with general predictions derived in Ref. [1].

On the other hand, a very sensitive probe of the phase transitional behavior is the ground-state energy with respect to the control parameters [11]

$$C(\lambda_i)\big|_{\{\lambda_j\}} = -\frac{\partial^2}{\partial \lambda_i^2} \varepsilon_0(\vec{\lambda}) \qquad , \qquad (11)$$

( all $\lambda_j's$ with $j \neq i$ are kept constant). In the above discussed thermodynamic analogy $\varepsilon_0(\vec{\lambda})$ is replaced by the equilibrium value of the thermodynamic potential $F_0(P,T)$. In our descriptions, by use of Eq. (7), ground-state energies are $A$, $A - \frac{B^2}{4C}$ for $\beta = 0$ and $\beta \neq 0$ respectively. From Eq. (11) the specific heats are

$$c^+(\beta_0 = 0) = 0 \quad \text{for} \quad x < 0.5 \quad \& \quad c^+(\beta_0 \neq 0) = \frac{12}{7}\text{N} \quad \text{for} \quad x > 0.5 \qquad (12)$$

These results propose any dependence of $C$ on $\hat{x}$ either above or below of $x_{critical} = 0.5$ and therefore, the values for the specific heat exponents are both zero. Also, this result suggests a discontinuity in the heat capacity in the phase transition point which in the agreement by Landau's theory [1,2]. We represented the behavior of specific heat in Figure 3 which one can find that, it has a jump at the critical point.

The classification of phase transitions that we follow in this paper and that is followed traditionally in the IBM is the Ehrenfest classification [12-15]. In Ehrenfest classification first order phase transitions appear when there exist a discontinuity in the first derivate of the energy with respect to the control parameter. Second order phase transitions appear when the second derivative of the energy with respect to the control parameter displays a discontinuity. It can be seen from Figure 4 that, first derive of the energy surface has a king at $x_{critical}$. This corresponds to a second order phase transition, as the second derivate is discontinuous.

In order to identify other critical exponents, according to the Landau theory, by use of Eq.(7), the potential energy surface becomes as

$$E(\beta) = A - h\beta + B\beta^2 + C\beta^4 + .... \qquad , \qquad (13)$$

Where, $h\beta$, represents the contribution of intensive parameter, $h$, for points off the $h = 0$ coexistence curve. The equilibrium equation of state is $(\partial E/\partial \beta)_{\hat{x},h} = 0$ which cause to (for any small $h$)

$$h = 2B\beta + 4C\beta^3 \qquad , \qquad (14)$$

On the other hand, it reduces to its former representation for $h = 0$. The susceptibility may be found as

$$X_x^{-1} = (\frac{\partial h}{\partial \beta})\bigg|_{\hat{x}} = 2N\hat{x} + \frac{7}{2}N\beta^2 \quad , \tag{15}$$

For $x < 0.5$ which we have $\beta = 0$ and consequently we get, $X_x^{-1} = 2N\hat{x}$, which gives the critical exponent equal to 1. For $x < 0.5$ with $h = 0$, Eq.(13) gives $\beta^2 = -\frac{12}{7}\hat{x}$ and therefore $X_x^{-1} = 4N(-\hat{x})$ or the critical exponent equal to 1. Along the critical isotherm, i.e. in the phase transition point, namely $\hat{x} = 0$, and $h = \frac{7}{6}N\beta^3$ which this means, critical exponent is equal to 3.

Our results, i.e. behavior of order parameter about critical point, discontinuity of the second order derivative of energy respect to order parameter, suggest a second order shape phase transition between *U(5)* and *O(6)* limits of IBM. Also, critical exponent and their capability to describe the order of quantum phase transition may be interpreted a new technique to explore shape phase transitions in complex systems.

3. Summery and conclusion

In this contribution, we show that, $U(5) \leftrightarrow O(6)$ shape phase transition are closely related to Landau theory of phase transition and explore some of the analogies with thermodynamics. Also, a detailed analysis of the critical exponents of ground state quantum phase transition is presented. We find that, critical exponents in two frameworks are similar. Based on a discontinuity in the heat capacity in the phase transition point, we can conclude the order of the phase transition.

Figures

Figure1. Energy surface of transitional Hamiltonian. Different panels describe dependence of energy surfaces on the order parameter, $\beta$, above and below of phase transition point, $x_c$.

Figure 2. Typical behavior of order parameter, $\beta$, at a second order phase transition.

Figure3. Equilibrium deformation, $\beta_0(x)$ for second order phase transition (a) and (b) specific heat of the ground state.

Figure4. Variation of energy surface and its first derivative respect to order parameter, it has a king at $x_{critical}$ which suggest a second order phase transition.

Figure 1.

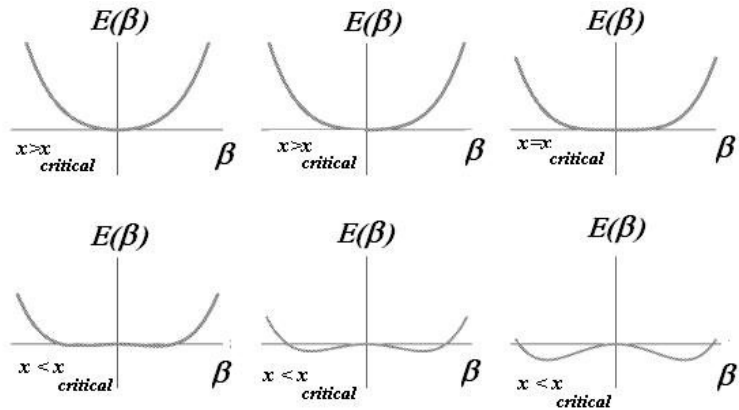

Figure 2.

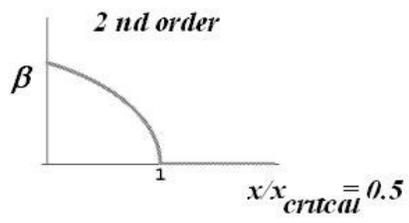

Figure 3.

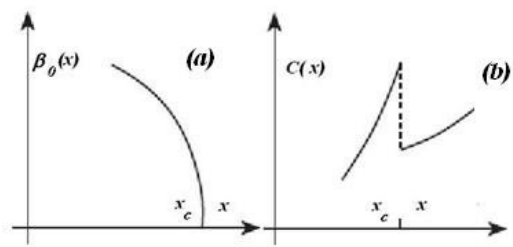

Figure 4.

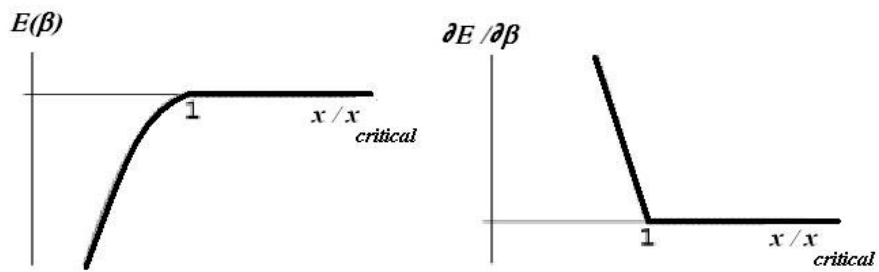